\documentclass[11pt]{article}
\usepackage{amssymb,amsmath,amsfonts}
\usepackage{graphicx}
\usepackage{graphics}
\usepackage{eepic,epsfig}

\begin{document}

\title{Casimir densities for two spherical branes in Rindler-like spacetimes}
\author{ A. A. Saharian$^{1,2}$\thanks{%
E-mail: saharyan@server.physdep.r.am } and M. R. Setare$^{3}$\thanks{%
E-mail: rezakord@ipm.ir} \\
\textit{$^1$ Department of Physics, Yerevan State University, Yerevan,
Armenia } \\
\textit{$^2$ Departamento de F\'{\i}sica-CCEN, Universidade Federal da Para%
\'{\i}ba, }\\
\textit{Jo\~{a}o Pessoa, PB, Brazil } \\
\textit{$^{3}$ Department of Science, Payame Noor University, Bijar, Iran}}
\maketitle

\begin{abstract}
Wightman function, the vacuum expectation values of the field square and the
energy-momentum tensor are evaluated for a scalar field obeying the Robin
boundary conditions on two spherical branes in $(D+1)$-dimensional
Rindler-like spacetime $Ri\times S^{D-1}$, with a two-dimensional Rindler
spacetime $Ri$. This spacetime approximates the near horizon geometry of $%
(D+1)$-dimensional black hole. By using the generalized Abel-Plana formula,
the vacuum expectation values are presented as the sum of single brane and
second brane induced parts. Various limiting cases are studied. The vacuum
forces acting on the branes are decomposed into the self-action and
interaction terms. The interaction forces are investigated as functions of
the brane locations and coefficients in the boundary conditions.
\end{abstract}

\bigskip

PACS number(s): 03.70.+k, 04.62.+v, 11.10.Kk

\section{Introduction}

\label{sec:Int}

Motivated by string/M theory, the AdS/CFT correspondence, and the hierarchy
problem of particle physics, braneworld models were studied actively in
recent years (for a review see \cite{Ruba01}). In these models, our world is
represented by a sub-manifold, a three-brane, embedded in a higher
dimensional spacetime. In particular, a well studied example is when the
bulk is an AdS space. The braneworld corresponds to a manifold with
boundaries and all fields which propagate in the bulk will give Casimir-type
contributions to the vacuum energy, and as a result to the vacuum forces
acting on the branes. In dependence of the type of a field and boundary
conditions imposed, these forces can either stabilize or destabilize the
braneworld. In addition, the Casimir energy gives a contribution to both the
brane and bulk cosmological constants and, hence, has to be taken into
account in the self-consistent formulation of the braneworld dynamics.
Motivated by these, the role of quantum effects in braneworld scenarios has
received a great deal of attention. For a conformally coupled scalar this
effect was initially studied in Ref. \cite{Fabi00} in the context of
M-theory, and subsequently in Refs. \cite{Noji00a} for a background
Randall-Sundrum geometry. The models with dS and AdS branes, and higher
dimensional brane models are considered as well \cite{Noji00b}.

In view of these recent developments, it seems interesting to generalize the
study of quantum effects to other types of bulk spacetimes. In particular,
it is of interest to consider non-Poincar\'{e} invariant braneworlds, both
to better understand the mechanism of localized gravity and for possible
cosmological applications. Bulk geometries generated by higher-dimensional
black holes are of special interest. In these models, the tension and the
position of the brane are tuned in terms of black hole mass and cosmological
constant and brane gravity trapping occurs in just the same way as in the
Randall-Sundrum model. Braneworlds in the background of the AdS black hole
were studied in \cite{AdSbhworld}. Like pure AdS space the AdS black hole
may be superstring vacuum. It is of interest to note that the phase
transitions which may be interpreted as confinement-deconfinement transition
in AdS/CFT setup may occur between pure AdS and AdS black hole \cite{Witt98}%
. Though, in the generic black hole background the investigation of
brane-induced quantum effects is technically complicated, the exact
analytical results can be obtained in the near horizon and large mass limit
when the brane is close to the black hole horizon. In this limit the black
hole geometry may be approximated by the Rindler-like manifold (for some
investigations of quantum effects on background of Rindler-like spacetimes
see \cite{Byts96} and references therein).

In the previous papers \cite{Saha05,Saha06} we have considered the vacuum
densities induced by a spherical brane in the bulk $Ri\times S^{D-1}$, where
$Ri$ is a two-dimensional Rindler spacetime. Continuing in this direction,
in the present paper we investigate the Wightman function, the vacuum
expectation values of the field square and the energy-momentum tensor for a
scalar field with an arbitrary curvature coupling parameter for two
spherical branes on the same bulk. Though the corresponding operators are
local, due to the global nature of the vacuum, these expectation values
describe the global properties of the bulk and carry an important
information about the physical structure of the quantum field at a given
point. The expectation value of the energy-momentum tensor acts as the
source of gravity in the Einstein equations and, hence, plays an important
role in modelling a self-consistent dynamics involving the gravitational
field. In addition to applications in braneworld models on the AdS black
hole bulk, the problem under consideration is also of separate interest as
an example with gravitational and boundary-induced polarizations of the
vacuum, where all calculations can be performed in a closed form. Note that
the vacuum densities induced by a single and two parallel flat branes in the
bulk geometry $Ri\times R^{D-1}$ for both scalar and electromagnetic fields
are investigated in \cite{Cand77,Saha06b}.

The paper is organized as follows. In the next section we consider the
positive frequency Wightman functions in the region between two branes. On
the basis of the generalized Abel-Plana formula, we present this function in
the form of the sum of single brane and second brane induced parts. By using
expression for the Wightman function, in section \ref{sec:VEVEMT} we
investigate the vacuum expectation values of the field square and the
energy-momentum tensor. Various limiting cases of the general formulae are
studied. In section \ref{sec:IntForce} the vacuum forces acting on the
branes due to the presence of the second brane are evaluated by making use
of the expression for the radial vacuum stress. The main results of the
paper are summarized in section \ref{sec:Conc}.

\section{Wightman function}

\label{sec:WF}

We consider a scalar field $\varphi (x)$ propagating on background of $(D+1)$%
-dimensional Rindler-like spacetime $Ri\times S^{D-1}$. The corresponding
metric is described by the line element%
\begin{equation}
ds^{2}=\xi ^{2}d\tau ^{2}-d\xi ^{2}-r_{H}^{2}d\Sigma _{D-1}^{2},
\label{ds22}
\end{equation}%
with the Rindler-like $(\tau ,\xi )$ part and $d\Sigma _{D-1}^{2}$ is the
line element for the space with positive constant curvature with the Ricci
scalar $R=(D-2)(D-1)/r_{H}^{2}$. Line element (\ref{ds22}) describes the
near horizon geometry of $(D+1)$-dimensional topological black hole with the
line element \cite{Mann97}%
\begin{equation}
ds^{2}=A_{H}(r)dt^{2}-\frac{dr^{2}}{A_{H}(r)}-r^{2}d\Sigma _{D-1}^{2},
\label{ds21}
\end{equation}%
where $A_{H}(r)=k+r^{2}/l^{2}-r_{0}^{D}/l^{2}r^{n}$, $n=D-2$, and
the parameter $k$ classifies the horizon topology, with $k=0,-1,1$
corresponding to flat, hyperbolic, and elliptic horizons,
respectively. The parameter $l$ is related to the bulk
cosmological constant and the parameter $r_{0}$ depends on the
mass of the black hole. In the non extremal case the function
$A_{H}(r)$ has a simple zero
at $r=r_{H}$, and in the near horizon limit, introducing new coordinates $%
\tau $ and $\rho $ in accordance with%
\begin{equation}
\tau =A_{H}^{\prime }(r_{H})t/2,\quad r-r_{H}=A_{H}^{\prime
}(r_{H})\xi ^{2}/4,  \label{tau}
\end{equation}%
the line element is written in the form (\ref{ds22}). Note that for a $(D+1)$%
-dimensional Schwarzschild black hole \cite{Call88} one has $%
A_{H}(r)=1-(r_{H}/r)^{n}$ and, hence, $A_{H}^{\prime }(r_{H})=n/r_{H}$.

The field equation is in the form%
\begin{equation}
\left( g^{ik}\nabla _{i}\nabla _{k}+m^{2}+\zeta R\right) \varphi (x)=0,
\label{fieldeq1}
\end{equation}%
where $\zeta $ is the curvature coupling parameter. Below we will assume
that the field satisfies the Robin boundary conditions on the hypersurfaces $%
\xi =a$ and $\xi =b$, $a<b$,
\begin{equation}
\left. \left( A_{j}+B_{j}\frac{\partial }{\partial \xi }\right) \varphi
\right\vert _{\xi =j}=0,\quad j=a,b,  \label{bound1}
\end{equation}%
with constant coefficients $A_{j}$ and $B_{j}$. The Dirichlet and Neumann
boundary conditions are obtained as special cases. In accordance with \ (\ref%
{tau}), the hypersurface $\xi =j$ corresponds to the spherical shell near
the black hole horizon with the radius $r_{j}=r_{H}+A_{H}^{\prime
}(r_{H})j^{2}/4$.

The branes divide the bulk into three regions corresponding to $0<\xi <a$, $%
a<\xi <b$, and $b<\xi <\infty $. In general, the coefficients in the
boundary conditions (\ref{bound1}) can be different for separate regions. In
the corresponding braneworld scenario based on the orbifolded version of the
model the region between the branes is employed only and the ratio $%
A_{j}/B_{j}$ for untwisted bulk scalars is related to the brane mass
parameters $c_{j}$\ of the field by the formula \cite{Saha05}
\begin{equation}
\frac{A_{j}}{B_{j}}=\frac{1}{2}\left( c_{j}-\frac{\zeta }{j}\right) ,\;j=a,b.
\label{ABbraneworld}
\end{equation}%
For a twisted scalar the Dirichlet boundary conditions are obtained on both
branes.

To evaluate the vacuum expectation values (VEVs) of the field square and the
energy-momentum tensor we need a complete set of eigenfunctions satisfying
the boundary conditions (\ref{bound1}). In accordance with the problem
symmetry, below we shall use the hyperspherical angular coordinates $%
(\vartheta ,\phi )=(\theta _{1},\theta _{2},\ldots ,\theta _{n},\phi )$ on $%
S^{D-1}$\ with $0\leqslant \theta _{k}\leqslant \pi $, $k=1,\ldots ,n$, and $%
0\leqslant \phi \leqslant 2\pi $. In these coordinates the eigenfunctions in
the region between the branes can be written in the form%
\begin{equation}
\varphi _{\alpha }(x)=C_{\alpha }Z_{i\omega }^{(b)}(\lambda _{l}\xi ,\lambda
_{l}b)Y(m_{k};\vartheta ,\phi )e^{-i\omega \tau },  \label{eigfunc1}
\end{equation}%
where $m_{k}=(m_{0}\equiv l,m_{1},\ldots m_{n})$, and $m_{1},m_{2},\ldots
m_{n}$ are integers such that $0\leqslant m_{n-1}\leqslant \cdots \leqslant
m_{1}\leqslant l$, $-m_{n-1}\leqslant m_{n}\leqslant m_{n-1}$. In Eq. (\ref%
{eigfunc1}) $Y(m_{k};\vartheta ,\phi )$ is the spherical harmonic of degree $%
l$ \cite{Erdelyi}, and
\begin{equation}
Z_{i\omega }^{(j)}(u,v)=\bar{I}_{i\omega }^{(j)}(v)K_{i\omega }(u)-\bar{K}%
_{i\omega }^{(j)}(v)I_{i\omega }(u),\;j=a,b,  \label{Deigfunc}
\end{equation}%
with $I_{i\omega }(x)$ and $K_{i\omega }(x)$ being the modified Bessel
functions with the imaginary order,
\begin{equation}
\quad \lambda _{l}=\frac{1}{r_{H}}\sqrt{l(l+n)+\zeta n(n+1)+m^{2}r_{H}^{2}}%
\,.  \label{lambdal}
\end{equation}%
Here and below for a given function $f(z)$ we use the barred notations
\begin{equation}
\bar{f}^{(j)}(z)=A_{j}f(z)+\frac{B_{j}}{j}zf^{\prime }(z),\quad j=a,b.
\label{fbarnot}
\end{equation}

Functions (\ref{eigfunc1}) satisfy the boundary condition on the brane $\xi
=b$. From the boundary condition on the brane $\xi =a$ we find that the
possible values for $\omega $ are roots to the equation
\begin{equation}
Z_{i\omega }(\lambda _{l}a,\lambda _{l}b)=0,  \label{Deigfreq}
\end{equation}%
with the notation
\begin{equation}
Z_{\omega }(u,v)=\bar{I}_{\omega }^{(b)}(v)\bar{K}_{\omega }^{(a)}(u)-\bar{K}%
_{\omega }^{(b)}(v)\bar{I}_{\omega }^{(a)}(u).  \label{Zomega}
\end{equation}%
For a fixed $\lambda _{l}$, the equation (\ref{Deigfreq}) has an infinite
set of real solutions with respect to $\omega $. We will denote them by $%
\omega _{n}=\omega _{n}(\lambda _{l}a,\lambda _{l}b)$, $\omega _{n}>0$, $%
n=1,2,\ldots $, and will assume that they are arranged in the ascending
order $\omega _{n}<\omega _{n+1}$. In addition to the real zeros, in
dependence of the values of the ratios $jA_{j}/B_{j}$, equation (\ref%
{Deigfreq}) can have a finite set of purely imaginary solutions. The
presence of such solutions leads to the modes with an imaginary frequency
and, hence, to the unstable vacuum. In the consideration below we will
assume the values of the coefficients in boundary conditions (\ref{bound1})
for which the imaginary solutions are absent and the vacuum is stable.

The coefficient $C_{\alpha }$ in (\ref{eigfunc1}) can be found from the
normalization condition%
\begin{equation}
r_{H}^{D-1}\int d\Omega \int_{a}^{b}\frac{d\xi }{\xi }\varphi _{\alpha }%
\overleftrightarrow{\partial }_{\tau }\varphi _{\alpha ^{\prime }}^{\ast
}=i\delta _{\alpha \alpha ^{\prime }}.  \label{normcond}
\end{equation}%
where the integration goes over the region between two spheres. Substituting
eigenfunctions (\ref{eigfunc1}) and using the relation $\int \left\vert
Y(m_{k};\vartheta ,\phi )\right\vert ^{2}d\Omega =N(m_{k})$ for spherical
harmonics, one finds%
\begin{equation}
C_{\alpha }^{2}=\left. \frac{r_{H}^{1-D}\bar{I}_{i\omega }^{(a)}(\lambda
_{l}a)}{N(m_{k})\bar{I}_{i\omega }^{(b)}(\lambda _{l}b)\frac{\partial }{%
\partial \omega }Z_{i\omega }(\lambda _{l}a,\lambda _{l}b)}\right\vert
_{\omega =\omega _{n}}.  \label{Calfa}
\end{equation}%
The explicit form for $N(m_{k})$ is given in \cite{Erdelyi} and will not be
necessary for the following considerations in this paper.

First of all we evaluate the positive frequency Wightman function%
\begin{equation}
G^{+}(x,x^{\prime })=\langle 0\vert \varphi (x)\varphi (x^{\prime })\vert
0\rangle ,  \label{W1}
\end{equation}%
where $|0\rangle $ is the amplitude for the corresponding vacuum state. The
VEVs for the field square and the energy-momentum tensor are obtained from
this function in the coincidence limit of the arguments. In addition, the
Wightman function determines the response of a particle detector in given
state of motion. By expanding the field operator over eigenfunctions and
using the commutation relations one can see that%
\begin{equation}
G^{+}(x,x^{\prime })=\sum_{\alpha }\varphi _{\alpha }(x)\varphi _{\alpha
}^{\ast }(x^{\prime }).  \label{W2}
\end{equation}%
Substituting eigenfunctions (\ref{eigfunc1}) into this mode sum formula and
by making use of the addition theorem%
\begin{equation}
\sum_{m_{k}}\frac{Y(m_{k};\vartheta ,\phi )}{N(m_{k})}Y(m_{k};\vartheta
^{\prime },\phi ^{\prime })=\frac{2l+n}{nS_{D}}C_{l}^{n/2}(\cos \theta ),
\label{addtheor}
\end{equation}%
for the Wightman function in the region between the branes one finds
\begin{eqnarray}
G^{+}(x,x^{\prime }) &=&\frac{r_{H}^{1-D}}{nS_{D}}\sum_{l=0}^{\infty
}(2l+n)C_{l}^{n/2}(\cos \theta )\sum_{n=1}^{\infty }\frac{\bar{I}_{i\omega
}^{(a)}(\lambda _{l}a)e^{-i\omega (\tau -\tau ^{\prime })}}{\bar{I}_{i\omega
}^{(b)}(\lambda _{l}b)\frac{\partial }{\partial \omega }Z_{i\omega }(\lambda
_{l}a,\lambda _{l}b)}  \notag \\
&&\times \left. Z_{i\omega }^{(b)}(\lambda _{l}\xi ,\lambda _{l}b)Z_{i\omega
}^{(b)}(\lambda _{l}\xi ^{\prime },\lambda _{l}b)\right\vert _{\omega
=\omega _{n}}.  \label{Wigh1}
\end{eqnarray}%
In this formula, $S_{D}=2\pi ^{D/2}/\Gamma (D/2)$ is the volume of the unit $%
(D-1)$-sphere, $C_{l}^{n/2}(x)$ is the Gegenbauer polynomial of degree $l$
and order $n/2$, $\theta $ is the angle between directions $(\vartheta ,\phi
)$ and $(\vartheta ^{\prime },\phi ^{\prime })$.

As the normal modes $\omega _{n}$ are not explicitly known and the terms
with large $n$ are highly oscillatory, the Wightman function in the form (%
\ref{Wigh1}) is not convenient. For the further evaluation we apply to the
sum over $n$ the summation formula derived in Ref. \cite{Saha06b} on the
basis of the generalized Abel-Plana formula \cite{SahRev}:%
\begin{equation}
\sum_{n=1}^{\infty }\frac{\bar{I}_{-i\omega _{n}}^{(b)}(v)\bar{I}_{i\omega
_{n}}^{(a)}(u)}{\frac{\partial }{\partial z}Z_{iz}(u,v)|_{z=\omega _{n}}}%
F(\omega _{n})=\int_{0}^{\infty }dz\frac{\sinh \pi z}{\pi ^{2}}%
\,F(z)-\int_{0}^{\infty }dz\frac{F(iz)+F(-iz)}{2\pi Z_{z}(u,v)}\bar{I}%
_{z}^{(a)}(u)\bar{I}_{-z}^{(b)}(v).  \label{Sumformula}
\end{equation}%
As a function $F(z)$ in this formula we choose
\begin{equation}
F(z)=\frac{Z_{iz}^{(b)}(\lambda _{l}\xi ,\lambda _{l}\lambda
b)Z_{iz}^{(b)}(\lambda _{l}\xi ^{\prime },\lambda _{l}b)}{\bar{I}%
_{iz}^{(b)}(\lambda _{l}b)\bar{I}_{-iz}^{(b)}(\lambda _{l}b)}e^{-iz(\tau
-\tau ^{\prime })}.  \label{FtoAPF}
\end{equation}%
The conditions for the formula (\ref{Sumformula}) to be valid are satisfied
if $a^{2}e^{|\tau -\tau ^{\prime }|}<\xi \xi ^{\prime }$. For the Wightman
function one obtains the expression%
\begin{eqnarray}
G^{+}(x,x^{\prime }) &=&G^{+}(x,x^{\prime };b)-\frac{r_{H}^{1-D}}{\pi nS_{D}}%
\sum_{l=0}^{\infty }(2l+n)C_{l}^{n/2}(\cos \theta )\int_{0}^{\infty }d\omega
\,\Omega _{b\omega }(\lambda _{l}a,\lambda _{l}b)  \notag \\
&&\times Z_{\omega }^{(b)}(\lambda _{l}\xi ,\lambda _{l}b)Z_{\omega
}^{(b)}(\lambda _{l}\xi ^{\prime },\lambda _{l}b)\cosh [\omega (\tau -\tau
^{\prime })],  \label{Wigh3}
\end{eqnarray}%
where
\begin{equation}
\Omega _{b\omega }(u,v)=\frac{\bar{I}_{\omega }^{(a)}(u)}{\bar{I}_{\omega
}^{(b)}(v)Z_{\omega }(u,v)}.  \label{Omega2}
\end{equation}%
In Eq. (\ref{Wigh3})
\begin{eqnarray}
G^{+}(x,x^{\prime };b) &=&\frac{r_{H}^{1-D}}{\pi ^{2}nS_{D}}%
\sum_{l=0}^{\infty }(2l+n)C_{l}^{n/2}(\cos \theta )\int_{0}^{\infty }d\omega
\sinh (\pi \omega )  \notag \\
&&\times e^{-i\omega (\tau -\tau ^{\prime })}\frac{Z_{i\omega
}^{(b)}(\lambda _{l}\xi ,\lambda b)Z_{i\omega }^{(b)}(\lambda _{l}\xi
^{\prime },\lambda _{l}b)}{\bar{I}_{i\omega }^{(b)}(\lambda _{l}b)\bar{I}%
_{-i\omega }^{(b)}(\lambda _{l}b)},  \label{Wigh1pl}
\end{eqnarray}%
is the Wightman function in the region $\xi <b$ for a single brane at $\xi
=b $ and the second term on the right is induced by the presence of the
brane at $\xi =a$. The function (\ref{Wigh3}) is investigated in Ref. \cite%
{Saha05} and can be presented in the form
\begin{equation}
G^{+}(x,x^{\prime };b)=G_{0}^{+}(x,x^{\prime })+\langle \varphi (x)\varphi
(x^{\prime })\rangle ^{(b)},  \label{G+2}
\end{equation}%
where $G_{0}^{+}(x,x^{\prime })$ is the Wightman function for the geometry
without boundaries and the part%
\begin{eqnarray}
\langle \varphi (x)\varphi (x^{\prime })\rangle ^{(b)} &=&-\frac{r_{H}^{1-D}%
}{\pi nS_{D}}\sum_{l=0}^{\infty }(2l+n)C_{l}^{n/2}(\cos \theta
)\int_{0}^{\infty }d\omega \frac{\bar{K}_{\omega }^{(b)}(\lambda _{l}b)}{%
\bar{I}_{\omega }^{(b)}(\lambda _{l}b)}  \notag \\
&&\times I_{\omega }(\lambda _{l}\xi )I_{\omega }(\lambda _{l}\xi ^{\prime
})\cosh [\omega (\tau -\tau ^{\prime })]  \label{phi212}
\end{eqnarray}%
is induced in the region $\xi <b$ by the presence of the brane at $\xi =b$.
Note that the representation (\ref{G+2}) with (\ref{phi212}) is valid under
the assumption $\xi \xi ^{\prime }<b^{2}e^{|\tau -\tau ^{\prime }|}$. As it
has been shown in \cite{Saha05}, the Wightman function for the boundary-free
geometry may be written in the form%
\begin{eqnarray}
G_{0}^{+}(x,x^{\prime }) &=&\tilde{G}_{0}^{+}(x,x^{\prime })-\frac{%
r_{H}^{1-D}}{\pi ^{2}nS_{D}}\sum_{l=0}^{\infty }(2l+n)C_{l}^{n/2}(\cos
\theta )  \notag \\
&&\times \int_{0}^{\infty }d\omega e^{-\omega \pi }\cos [\omega (\tau -\tau
^{\prime })]K_{i\omega }(\lambda _{l}\xi )K_{i\omega }(\lambda _{l}\xi
^{\prime }),  \label{GM1}
\end{eqnarray}%
where $\tilde{G}_{0}^{+}(x,x^{\prime })$ is the Wightman function for the
bulk geometry $R^{2}\times S^{D-1}$. Outside the horizon the divergences in
the coincidence limit of the expression on the right of (\ref{GM1}) are
contained in the first term.

It can be seen that the Wightman function in the region between
the branes can be also presented in the form
\begin{eqnarray}
G^{+}(x,x^{\prime }) &=&G^{+}(x,x^{\prime };a)-\frac{r_{H}^{1-D}}{\pi nS_{D}}%
\sum_{l=0}^{\infty }(2l+n)C_{l}^{n/2}(\cos \theta )\int_{0}^{\infty }d\omega
\,\Omega _{a\omega }(\lambda _{l}a,\lambda _{l}b)  \notag \\
&&\times Z_{\omega }^{(a)}(\lambda _{l}\xi ,\lambda _{l}a)Z_{\omega
}^{(a)}(\lambda _{l}\xi ^{\prime },\lambda _{l}a)\cosh [\omega (\tau -\tau
^{\prime })],  \label{Wigh31}
\end{eqnarray}%
with the notation
\begin{equation}
\Omega _{a\omega }(u,v)=\frac{\bar{K}_{\omega }^{(b)}(v)}{\bar{K}_{\omega
}^{(a)}(u)Z_{\omega }(u,v)}.  \label{Oma}
\end{equation}
In this representation,%
\begin{equation}
G^{+}(x,x^{\prime };a)=G_{0}^{+}(x,x^{\prime })+\langle \varphi (x)\varphi
(x^{\prime })\rangle ^{(a)}  \label{G+1}
\end{equation}%
is the Wightman function in the region $\xi >a$ for a single brane at $\xi
=a $, and
\begin{eqnarray}
\langle \varphi (x)\varphi (x^{\prime })\rangle ^{(a)} &=&-\frac{r_{H}^{1-D}%
}{\pi nS_{D}}\sum_{l=0}^{\infty }(2l+n)C_{l}^{n/2}(\cos \theta
)\int_{0}^{\infty }d\omega \frac{\bar{I}_{\omega }^{(a)}(\lambda _{l}a)}{%
\bar{K}_{\omega }^{(a)}(\lambda _{l}a)}  \notag \\
&&\times K_{\omega }(\lambda _{l}\xi )K_{\omega }(\lambda _{l}\xi ^{\prime
})\cosh [\omega (\tau -\tau ^{\prime })].  \label{phi211}
\end{eqnarray}%
Two representations of the Wightman function, given by Eqs. (\ref{Wigh3})
and (\ref{Wigh31}), are obtained from each other by the replacements%
\begin{equation}
a\rightleftarrows b,\quad I_{\omega }\rightleftarrows K_{\omega }.
\label{replacement}
\end{equation}%
In the coincidence limit the second term on the right of formula (\ref{Wigh3}%
) is finite on the brane $\xi =b$ and diverges on the brane at $\xi =a$,
whereas the second term on the right of Eq. (\ref{Wigh31}) is finite on the
brane $\xi =a$ and is divergent for $\xi =b$. Consequently, the forms (\ref%
{Wigh3}) and (\ref{Wigh31}) are convenient for the investigations of the
VEVs near the branes $\xi =b$ and $\xi =a$, respectively.

We have investigated the Whightman function in the region between two branes
for an arbitrary ratio of boundary coefficients $A_{j}/B_{j}$. Note that in
the orbifolded version of the model the integration in the normalization
integral goes over two copies of the bulk manifold. This leads to the
additional coefficient $1/2$ in the expression (\ref{Calfa}) for the
normalization coefficient $C_{\alpha }$. Hence, the Whightman function in
the orbifolded braneworld case is given by formula (\ref{Wigh3}) with an
additional factor $1/2$ in the second term on the right and in formula (\ref%
{Wigh1pl}). As it has been mentioned above this function corresponds to the
braneworld in the AdS black hole bulk in the limit when the branes are close
to the black hole horizon.

\section{Casimir densities}

\label{sec:VEVEMT}

\subsection{VEV for the field square}

In this section we will consider the VEVs of the field square and the
energy-momentum tensor in the region between the branes. In the coincidence
limit, taking into account the relation $C_{l}^{n/2}(1)=\Gamma (l+n)/\Gamma
(n)l!$, from the formulae for the Wightman function one obtains two
equivalent forms for the VEV\ of the field square:%
\begin{eqnarray}
\langle 0\vert \varphi ^{2}\vert 0\rangle &=&\langle 0_{0}\vert \varphi
^{2}\vert 0_{0}\rangle +\langle \varphi ^{2}\rangle ^{(j)}  \notag \\
&&-\frac{r_{H}^{1-D}}{\pi S_{D}}\sum_{l=0}^{\infty }D_{l}\int_{0}^{\infty
}d\omega \,\Omega _{j\omega }(\lambda _{l}a,\lambda _{l}b)Z_{\omega
}^{(j)2}(\lambda _{l}\xi ,\lambda _{l}j),  \label{phi2sq1}
\end{eqnarray}%
corresponding to $j=a$ and $j=b$, and $|0_{0}\rangle $ is the amplitude for
the vacuum without boundaries,
\begin{equation}
D_{l}=(2l+D-2)\frac{\Gamma (l+D-2)}{\Gamma (D-1)l!}  \label{Dl}
\end{equation}%
is the degeneracy of each angular mode with given $l$. The VEV $\langle
0_{0}\vert \varphi ^{2}\vert 0_{0}\rangle $ is obtained from the
corresponding Wightman function given by (\ref{GM1}). For the points outside
the horizon, the renormalization procedure is needed for the first term on
the right only, which corresponds to the VEV in the geometry $R^{2}\times
S^{D-1}$. This procedure is realized in \cite{Saha05} on the base of the
zeta function technique.

In Eq. (\ref{phi2sq1}), the part $\langle \varphi ^{2}\rangle ^{(j)}$ is
induced by a single brane at $\xi =j$ when the second brane is absent. For
the geometry of a single brane at $\xi =a$, from (\ref{phi211}) one has
\begin{equation}
\langle \varphi ^{2}\rangle ^{(a)}=-\frac{r_{H}^{1-D}}{\pi S_{D}}%
\sum_{l=0}^{\infty }D_{l}\int_{0}^{\infty }d\omega \frac{\bar{I}_{\omega
}^{(a)}(\lambda _{l}a)}{\bar{K}_{\omega }^{(a)}(\lambda _{l}a)}K_{\omega
}^{2}(\lambda _{l}\xi ).  \label{phi21plb}
\end{equation}%
The expression for $\langle \varphi ^{2}\rangle ^{(b)}$ is obtained from (%
\ref{phi21plb}) by replacements (\ref{replacement}) and is investigated in
\cite{Saha05}. The last term on the right of formula (\ref{phi2sq1}) is
induced by the presence of the second brane. It is finite on the brane at $%
\xi =j$ and diverges for the points on the other brane. By taking into
account the relation $Z_{\omega }^{(j)}(u,u)=B_{j}/j$, we see that for the
Dirichlet boundary condition this term vanishes on the brane $\xi =j$.

Let us consider the behavior of the single brane part (\ref{phi21plb}) in
asymptotic regions of the parameters. In the limit $\xi \rightarrow a$ this
part diverges and, hence, for points near the brane the main contribution
comes from large values $l$. By making use of the corresponding uniform
asymptotic expansions for the modified Bessel functions, to the leading
order we find
\begin{equation}
\langle \varphi ^{2}\rangle ^{(a)}\approx -\frac{\delta _{B_{a}}\Gamma
\left( \frac{D-1}{2}\right) }{(4\pi )^{\frac{D+1}{2}}(\xi -a)^{D-1}},
\label{phi2anear}
\end{equation}%
where $\delta _{B_{a}}=1$ for $B_{a}=0$ and $\delta _{B_{a}}=-1$ for $%
B_{a}\neq 0$. Hence, near the brane the brane-induced part is negative for
the Dirichlet boundary condition and is positive for non-Dirichlet boundary
condition. At large distances from the brane, $\xi \gg r_{H}$, the dominant
contribution into (\ref{phi21plb}) comes from the $l=0$ term and in the
leading order we have%
\begin{equation}
\langle \varphi ^{2}\rangle ^{(a)}\approx -\frac{r_{H}^{1-D}e^{-2\lambda
_{0}\xi }}{2S_{D}\lambda _{0}\xi }\int_{0}^{\infty }d\omega \frac{\bar{I}%
_{\omega }^{(a)}(\lambda _{0}a)}{\bar{K}_{\omega }^{(a)}(\lambda _{0}a)}.
\label{phi2afar}
\end{equation}%
In the limit when the position of the brane tends to the horizon, $%
a\rightarrow 0$, with fixed $\xi $, we use the formulae for the modified
Bessel functions with small values of the argument. The main contribution
into the integral comes from the lower limit of the integration and we
obtain the formula%
\begin{equation}
\langle \varphi ^{2}\rangle ^{(a)}\approx -\frac{r_{H}^{1-D}\delta _{B_{a}}}{%
2\pi S_{D}\ln ^{2}(r_{H}/a)}\sum_{l=0}^{\infty }D_{l}K_{0}^{2}(\lambda
_{l}\xi ).  \label{phi2anearhoriz}
\end{equation}

In the limit $r_{H}\rightarrow 0$ the curvature for the background spacetime
is large. In this limit $\lambda _{l}$ is also large. The exception is the
term $l=0$ for a minimally coupled scalar field for which $\lambda _{0}=m$.
For large values $\lambda _{l}$ the main contribution into the integral in (%
\ref{phi21plb}) comes from large values $\omega $. Introducing a new
integration variable $x=\omega /\lambda _{l}a$, we estimate the integral by
the Laplace method. This leads to the following result%
\begin{equation}
\langle \varphi ^{2}\rangle ^{(a)}\approx -\frac{\delta _{B_{a}}r_{H}^{1-D}a%
}{4\sqrt{\pi }S_{D}\xi }\frac{\exp [-2\sqrt{\zeta n(n+1)}(\xi -a)/r_{H}]}{%
\sqrt{\lambda _{0}(\xi -a)}}.  \label{phi2alargecurv}
\end{equation}%
For a minimally coupled scalar field the contribution of the terms with $%
l\geqslant 1$ is suppressed by the factor $e^{-2\lambda _{l}(\xi -a)}$ and
the dominant contribution comes from the $l=0$\ term:%
\begin{equation}
\langle \varphi ^{2}\rangle ^{(a)}=-\frac{r_{H}^{1-D}}{\pi S_{D}}%
\int_{0}^{\infty }d\omega \frac{\bar{I}_{\omega }^{(a)}(ma)}{\bar{K}_{\omega
}^{(a)}(ma)}K_{\omega }^{2}(m\xi ).  \label{phi2asmallrh}
\end{equation}%
As we see, the behavior of the VEV in the high curvature regime is
essentially different for minimally and non-minimally coupled
fields.

In the near-horizon limit, $a,\xi \ll r_{H}$, the main contribution into the
sum over $l$ in (\ref{phi21plb}) comes from large values $l$ corresponding
to $l\lesssim r_{H}/(\xi -a)$. To the leading order we can replace the
summation over $l$ by the integration in accordance with the formula%
\begin{equation}
\sum_{l=0}^{\infty }D_{l}f(\lambda _{l})\rightarrow \frac{2r_{H}^{D-1}}{%
\Gamma (D-1)}\int_{0}^{\infty }dk\,k^{D-2}f(\sqrt{k^{2}+m^{2}}).
\label{sumtoint}
\end{equation}%
Now it is easily seen that from (\ref{phi21plb}) we obtain the corresponding
result for the plate uniformly accelerated through the Fulling-Rindler
vacuum.

In the geometry of two branes, extracting the contribution from the second
brane, we can write the expression (\ref{phi2sq1}) for the VEV in the
symmetric form
\begin{equation}
\langle 0\left\vert \varphi ^{2}\right\vert 0\rangle =\langle
0_{0}\left\vert \varphi ^{2}\right\vert 0_{0}\rangle +\sum_{j=a,b}\langle
\varphi ^{2}\rangle ^{(j)}+\langle \varphi ^{2}\rangle ^{(ab)},
\label{phi2sq2n}
\end{equation}%
with the interference part%
\begin{equation}
\langle \varphi ^{2}\rangle ^{(ab)}=-\frac{r_{H}^{1-D}}{\pi S_{D}}%
\sum_{l=0}^{\infty }D_{l}\int_{0}^{\infty }d\omega \bar{I}_{\omega
}^{(a)}(\lambda _{l}a)\left[ \frac{Z_{\omega }^{(b)2}(\lambda _{l}\xi
,\lambda _{l}b)}{\bar{I}_{\omega }^{(b)}(\lambda _{l}b)Z_{\omega }(\lambda
_{l}a,\lambda _{l}b)}-\frac{K_{\omega }^{2}(\lambda _{l}\xi )}{\bar{K}%
_{\omega }^{(a)}(\lambda _{l}a)}\right] .  \label{phi2int}
\end{equation}%
An equivalent form for this part is obtained with the replacements (\ref%
{replacement}) in the integrand. The interference term
(\ref{phi2int}) is finite for all values of $\xi $ in the range
$a\leqslant \xi \leqslant b$, including the points on the branes.
The surface divergences are contained in the single brane parts
only.

Let us consider the behavior of the interference part in the VEV of the
field square in limiting regions for values of the parameters. First of all,
it can be seen that in the limit $a\rightarrow b$, to the leading order the
result for the parallel plates in the Minkowski bulk is obtained. When the
left brane tends to the horizon, $a\rightarrow 0$, the dominant contribution
comes from the lower limit of the integration in (\ref{phi2int}), and we
have
\begin{equation}
\langle \varphi ^{2}\rangle ^{(ab)}\approx \frac{r_{H}^{1-D}\delta _{B_{a}}}{%
2\pi S_{D}\ln ^{2}(r_{H}/a)}\sum_{l=0}^{\infty }D_{l}\frac{\bar{K}%
_{0}^{(b)}(\lambda _{l}b)}{\bar{I}_{0}^{(b)}(\lambda _{l}b)}\left[
2K_{0}(\lambda _{l}\xi )-\frac{\bar{K}_{0}^{(b)}(\lambda _{l}b)}{\bar{I}%
_{0}^{(b)}(\lambda _{l}b)}I_{0}(\lambda _{l}\xi )\right] I_{0}(\lambda
_{l}\xi ).  \label{phi2abnearhoriz}
\end{equation}%
In the limit $b\rightarrow \infty $ for fixed values $a$ and $\xi $, the
main contribution comes from the lowest mode $l=0$, and to the leading order
one finds%
\begin{equation}
\langle \varphi ^{2}\rangle ^{(ab)}\approx \frac{e^{-2\lambda _{0}b}}{%
S_{D}r_{H}^{D-1}}\frac{A_{b}-B_{b}\lambda _{0}}{A_{b}+B_{b}\lambda _{0}}%
\int_{0}^{\infty }d\omega \frac{\bar{I}_{\omega }^{(a)}(\lambda _{0}a)}{\bar{%
K}_{\omega }^{(a)}(\lambda _{0}a)}\left[ 2I_{\omega }(\lambda _{0}\xi )-%
\frac{\bar{I}_{\omega }^{(a)}(\lambda _{0}a)}{\bar{K}_{\omega
}^{(a)}(\lambda _{0}a)}K_{\omega }(\lambda _{0}\xi )\right] K_{\omega
}(\lambda _{0}\xi ),  \label{phi2abfar}
\end{equation}%
with the exponentially suppressed interference part. The behavior
of the interference part in the limit $r_{H}\rightarrow 0$ can be
investigated in the way similar to that for a single brane part.
For a non-minimally coupled scalar field the interference part is
dominated by the $l=0$ term and is suppressed by the factor $\exp
[-2\sqrt{\zeta n(n+1)}(b-a)/r_{H}]$. For a minimally coupled
scalar field the leading term is given by the $l=0$
summand with $\lambda _{0}=m$ and the interference part behaves as $%
r_{H}^{1-D}$. In the near-horizon limit, $a,b\ll r_{H}$, replacing the
summation by the integration in accordance with formula (\ref{sumtoint}), it
can be seen that from (\ref{phi2int}) the result for the geometry of two
parallel plates uniformly accelerated through the Fulling-Rindler vacuum is
obtained.

\subsection{Energy-momentum tensor}

The VEV of the energy-momentum tensor is expressed in terms of the Wightman
function as
\begin{equation}
\langle 0\vert T_{ik}\vert 0\rangle =\lim_{x^{\prime }\rightarrow x}\partial
_{i}\partial _{k}^{\prime }G^{+}(x,x^{\prime })+\left[ \left( \zeta -\frac{1%
}{4}\right) g_{ik}\nabla _{l}\nabla ^{l}-\zeta \nabla _{i}\nabla _{k}-\zeta
R_{ik}\right] \langle 0\vert \varphi ^{2}\vert 0\rangle ,  \label{vevemtW}
\end{equation}%
where $R_{ik}$ is the Ricci tensor for the bulk geometry. Making use of the
formulae for the Wightman function and the VEV of the field square, one
obtains two equivalent forms, corresponding to $j=a$ and $j=b$ (no summation
over $i$):
\begin{eqnarray}
\langle 0|T_{i}^{k}|0\rangle &=&\langle 0_{0}|T_{i}^{k}|0_{0}\rangle
+\langle T_{i}^{k}\rangle ^{(j)}-\delta _{i}^{k}\frac{r_{H}^{1-D}}{\pi S_{D}}%
\sum_{l=0}^{\infty }D_{l}\lambda _{l}^{2}  \notag \\
&&\times \int_{0}^{\infty }d\omega \,\Omega _{j\omega }(\lambda
_{l}a,\lambda _{l}b)F^{(i)}\left[ Z_{\omega }^{(j)}(\lambda _{l}\xi ,\lambda
_{l}j)\right] .  \label{Tik1}
\end{eqnarray}
In this formula, for a given function $g(z)$ we use the notations
\begin{eqnarray}
F^{(0)}[g(z)] &=&\left( \frac{1}{2}-2\zeta \right) \left[ \left( \frac{dg(z)%
}{dz}\right) ^{2}+\left( 1+\frac{\omega ^{2}}{z^{2}}\right) g^{2}(z)\right] +%
\frac{\zeta }{z}\frac{d}{dz}g^{2}(z)-\frac{\omega ^{2}}{z^{2}}g^{2}(z),
\label{f0} \\
F^{(1)}[g(z)] &=&-\frac{1}{2}\left( \frac{dg(z)}{dz}\right) ^{2}-\frac{\zeta
}{z}\frac{d}{dz}g^{2}(z)+\frac{1}{2}\left( 1+\frac{\omega ^{2}}{z^{2}}%
\right) g^{2}(z),  \label{f1} \\
F^{(i)}[g(z)] &=&\left( \frac{1}{2}-2\zeta \right) \left[ \left( \frac{dg(z)%
}{dz}\right) ^{2}+\left( 1+\frac{\omega ^{2}}{z^{2}}\right) g^{2}(z)\right] -%
\frac{g^{2}(z)}{D-1}\frac{\lambda _{l}^{2}-m^{2}}{\lambda _{l}^{2}},
\label{f23}
\end{eqnarray}%
with $g(z)=Z_{\omega }^{(j)}(z,\lambda _{l}j)$, where $i=2,\ldots ,D$ and
the indices 0,1 correspond to the coordinates $\tau $, $\xi $, respectively.
In formula (\ref{Tik1}),
\begin{equation}
\langle 0_{0}|T_{i}^{k}|0_{0}\rangle =\delta _{i}^{k}\frac{r_{H}^{1-D}}{\pi
^{2}S_{D}}\sum_{l=0}^{\infty }D_{l}\lambda _{l}^{2}\int_{0}^{\infty }d\omega
\sinh \pi \omega \,f^{(i)}[K_{i\omega }(\lambda _{l}\xi )]  \label{DFR}
\end{equation}%
is the corresponding VEV for the vacuum without boundaries, and the term $%
\langle T_{i}^{k}\rangle ^{(j)}$ is induced by the presence of a single
spherical brane located at $\xi =j$. For the brane at $\xi =a$ and in the
region $\xi >a$ one has (no summation over $i$)
\begin{equation}
\langle T_{i}^{k}\rangle ^{(a)}=-\delta _{i}^{k}\frac{r_{H}^{1-D}}{\pi S_{D}}%
\sum_{l=0}^{\infty }D_{l}\lambda _{l}^{2}\int_{0}^{\infty }d\omega \frac{%
\bar{I}_{\omega }^{(a)}(\lambda _{l}a)}{\bar{K}_{\omega }^{(a)}(\lambda
_{l}a)}F^{(i)}[K_{\omega }(\lambda _{l}\xi )].  \label{D1plateboundb}
\end{equation}%
For the geometry of a single brane at $\xi =b$, the corresponding expression
in the region $\xi <b$ is obtained from (\ref{D1plateboundb}) by the
replacements (\ref{replacement}). The expressions for the functions $%
f^{(i)}[g(z)]$ in (\ref{DFR}) are obtained from the corresponding
expressions for $F^{(i)}[g(z)]$ by the replacement $\omega \rightarrow
i\omega $. It can be easily seen that for a conformally coupled massless
scalar the boundary induced part in the energy-momentum tensor is traceless.
The boundary-free part (\ref{DFR}) and the single brane part $\langle
T_{i}^{k}\rangle ^{(j)}$ in the region $\xi <j$ are investigated in Ref.
\cite{Saha05}.

Now we turn to the investigation of the brane-induced VEVs in limiting
cases. First of all let us consider single brane part (\ref{D1plateboundb}).
At large distances from the brane, $\xi \gg r_{H}$, the main contribution
comes from the $l=0$ term and one has
\begin{equation}
\langle T_{i}^{k}\rangle ^{(a)}\approx \lambda _{0}^{2}\delta
_{i}^{k}F_{0}^{(i)}\langle \varphi ^{2}\rangle ^{(a)},  \label{Tikafar}
\end{equation}%
where $\langle \varphi ^{2}\rangle ^{(a)}$ is given by (\ref{phi2afar}) and%
\begin{equation}
F_{0}^{(0)}=1-4\zeta ,\;F_{0}^{(1)}=\frac{4\zeta -1}{2\lambda _{0}\xi }%
,\;F_{0}^{(2)}=1-4\zeta -\frac{1-m^{2}/\lambda _{0}^{2}}{D-1}.
\label{F0ifar}
\end{equation}%
In this limit the radial vacuum stress is suppressed by the factor $\lambda
_{0}\xi $ with respect to the corresponding energy density and azimuthal
stresses. In the limit $a\rightarrow 0$ when $\xi $ is fixed the main
contribution into the $\omega $-integral comes from the lower limit and to
the leading order we obtain%
\begin{equation}
\langle T_{i}^{k}\rangle ^{(a)}\approx -\frac{\delta
_{i}^{k}r_{H}^{1-D}\delta _{B_{a}}}{2\pi S_{D}\ln ^{2}(r_{H}/a)}%
\sum_{l=0}^{\infty }D_{l}\lambda _{l}^{2}F^{(i)}[K_{\omega }(\lambda _{l}\xi
)]_{\omega =0}.  \label{Tikanearhor}
\end{equation}%
For $r_{H}\rightarrow 0$, in the way similar to that for the field square it
can be seen that for a non-minimally coupled scalar field $\langle
T_{i}^{k}\rangle ^{(a)}$ is suppressed by the factor $\exp [-2\sqrt{\zeta
n(n+1)}(\xi -a)/r_{H}]$. For a minimally coupled scalar the main
contribution comes from the $l=0$ term and the brane induced VEV (\ref%
{D1plateboundb}) behaves like $r_{H}^{1-D}$.

Now let us present the VEV (\ref{Tik1}) in the form%
\begin{equation}
\langle 0|T_{i}^{k}|0\rangle =\langle 0_{0}|T_{i}^{k}|0_{0}\rangle
+\sum_{j=a,b}\langle T_{i}^{k}\rangle ^{(j)}+\langle T_{i}^{k}\rangle
^{(ab)},  \label{Tikdecomp}
\end{equation}%
where the interference part is given by the formula (no summation over $i$)
\begin{eqnarray}
\langle T_{i}^{k}\rangle ^{(ab)} &=&-\delta _{i}^{k}\frac{r_{H}^{1-D}}{\pi
S_{D}}\sum_{l=0}^{\infty }D_{l}\lambda _{l}^{2}\int_{0}^{\infty }d\omega
\bar{I}_{\omega }^{(a)}(\lambda _{l}a)  \notag \\
&&\times \left[ \frac{F^{(i)}[Z_{\omega }^{(b)}(\lambda _{l}\xi ,\lambda
_{l}b)]}{\bar{I}_{\omega }^{(b)}(\lambda _{l}b)Z_{\omega }(\lambda
_{l}a,\lambda _{l}b)}-\frac{F^{(i)}[K_{\omega }(\lambda _{l}\xi )]}{\bar{K}%
_{\omega }^{(a)}(\lambda _{l}a)}\right] .  \label{intterm1}
\end{eqnarray}%
The surface divergences are contained in the single brane parts and the term
(\ref{intterm1}) is finite for all values $a\leqslant \xi \leqslant b$. An
equivalent formula for $\langle T_{i}^{k}\rangle ^{(ab)}$ is obtained from
Eq. (\ref{intterm1}) by replacements (\ref{replacement}). The behavior of
the interference part (\ref{intterm1}) in the limits $a\rightarrow 0$ and $%
b\rightarrow \infty $ is similar to that for the field square. In the
near-horizon limit, $a,b\ll r_{H}$, for both single brane and interference
parts replacing the summation by the integration in accordance with formula (%
\ref{sumtoint}), the result for the parallel plates uniformly accelerated
through the Fulling-Rindler vacuum is obtained.

\section{Vacuum interaction forces between the branes}

\label{sec:IntForce}

In this section we will consider the vacuum forces acting on the branes. The
force acting per unit surface of the brane at $\xi =j$ is determined by the
radial component of the vacuum energy-momentum tensor evaluated at this
point. By using the decomposition of the VEV for the energy-momentum tensor
given by (\ref{Tik1}), the corresponding effective pressures, $%
p^{(j)}=-\langle T_{1}^{1}\rangle _{\xi =j}$, can be presented as the sum
\begin{equation}
p^{(j)}=p_{1}^{(j)}+p_{\mathrm{(int)}}^{(j)},\quad j=a,b,  \label{FintD}
\end{equation}%
where the first term on the right is the pressure for a single brane at $\xi
=j$ when the second brane is absent. This term is divergent due to the
surface divergences in the subtracted vacuum expectation values and needs
additional renormalization. This can be done, for example, by applying the
generalized zeta function technique to the corresponding mode sum. This
procedure is similar to that used in Ref. \cite{Saha06} for the evaluation
of the surface energy for a single brane. The second term on the right of
Eq. (\ref{FintD}), $p_{\mathrm{(int)}}^{(j)}$, is the pressure induced by
the presence of the second brane, and can be termed as an interaction force.
This term determines the force by which the scalar vacuum acts on the brane
due to the modification of the spectrum for the zero-point fluctuations by
the presence of the second brane. It is finite for all nonzero distances
between the branes and is not affected by the renormalization procedure.

For the brane at $\xi =j$ the interaction term is due to the third summand
on the right of Eq. (\ref{Tik1}). Substituting into this term $i=k=1$, $\xi
=j$ and using the Wronskian relation for the modified Bessel functions one
finds
\begin{equation}
p_{\mathrm{(int)}}^{(j)}=\frac{A_{j}^{2}}{2j^{2}}\frac{r_{H}^{1-D}}{\pi S_{D}%
}\sum_{l=0}^{\infty }D_{l}\int_{0}^{\infty }d\omega \left[ \left( \lambda
_{l}^{2}j^{2}+\omega ^{2}\right) \beta _{j}^{2}+4\zeta \beta _{j}-1\right]
\,\Omega _{j\omega }(\lambda _{l}a,\lambda _{l}b),  \label{pint2}
\end{equation}%
with $\beta _{j}=B_{j}/jA_{j}$. The interaction force acts on the surface $%
\xi =a+0$ for the brane at $\xi =a$ and on the surface $\xi =b-0$ for the
brane at $\xi =b$. In dependence of the values for the coefficients in the
boundary conditions, the effective pressures (\ref{pint2}) can be either
positive or negative, leading to repulsive or attractive forces,
respectively. For Dirichlet or Neumann boundary conditions on both branes
the interaction forces are always attractive. For Dirichlet boundary
condition on one brane and Neumann boundary condition on the other one has $%
p_{\mathrm{(int)}}^{(j)}>0$ and the interaction forces are repulsive for all
distances between the branes. Note that the interaction forces can also be
written in another equivalent form
\begin{eqnarray}
p_{\mathrm{(int)}}^{(j)} &=&\frac{n^{(j)}}{2j}\frac{r_{H}^{1-D}}{\pi S_{D}}%
\sum_{l=0}^{\infty }D_{l}\int_{0}^{\infty }d\omega \left[ 1+\frac{\left(
4\zeta -1\right) \beta _{j}}{\left( \lambda _{l}^{2}j^{2}+\omega ^{2}\right)
\beta _{j}^{2}+\beta _{j}-1}\right]  \notag \\
&&\times \frac{\partial }{\partial j}\ln \left\vert 1-\frac{\bar{I}_{\omega
}^{(a)}(\lambda _{l}a)\bar{K}_{\omega }^{(b)}(\lambda _{l}b)}{\bar{I}%
_{\omega }^{(b)}(\lambda _{l}b)\bar{K}_{\omega }^{(a)}(\lambda _{l}a)}%
\right\vert .  \label{pint3}
\end{eqnarray}

Now we turn to the investigation of the interaction forces in the asymptotic
regions of the parameters. In the limit $a\rightarrow b$ the dominant
contribution into the expression on the right of (\ref{pint2}) comes from
large values $l$ and $\omega $. Replacing the summation over $l$ by the
integration in accordance with $\sum_{l=0}^{\infty }D_{l}f(l)\rightarrow
2\int_{0}^{\infty }dl\,l^{D-2}f(l)/\Gamma (D-1)$, and using the uniform
asymptotic expansions for the modified Bessel functions, to the leading
order we find%
\begin{equation}
p_{\mathrm{(int)}}^{(j)}\approx \sigma _{ab}\frac{\Gamma \left( \frac{D+1}{2}%
\right) \zeta _{\mathrm{R}}(D+1)}{(4\pi )^{(D+1)/2}(b-a)^{D+1}},
\label{pjintnear}
\end{equation}%
where $\zeta _{\mathrm{R}}(x)$ is the Riemann zeta function, $\sigma
_{ab}=-1 $ for $\delta _{B_{a}}\delta _{B_{b}}=1$ and $\sigma _{ab}=1-2^{-D}$
for $\delta _{B_{a}}\delta _{B_{b}}=-1$. Hence, for small distances between
the branes the interaction forces are repulsive for the Dirichlet boundary
condition on one brane and non-Dirichlet boundary condition on the other and
are attractive for all other cases. Note that in the limit $a\rightarrow b$
the interaction part of the total vacuum force acting on the brane diverges,
whereas the renormalized single brane parts remain finite. From here it
follows that at small distances between the branes the interaction part
dominates.

When the left brane tends to the horizon, $a\rightarrow 0$, the main
contribution into the vacuum interaction forces comes from the lower limit
of the $\omega $-integral and one has%
\begin{eqnarray}
p_{\mathrm{(int)}}^{(a)} &\approx &-\frac{\delta _{B_{a}}r_{H}^{1-D}}{2\pi
S_{D}a^{2}\ln ^{3}(r_{H}/a)}\sum_{l=0}^{\infty }D_{l}\frac{\bar{K}%
_{0}^{(b)}(\lambda _{l}b)}{\bar{I}_{0}^{(b)}(\lambda _{l}b)},
\label{panearhor} \\
p_{\mathrm{(int)}}^{(b)} &\approx &\frac{r_{H}^{1-D}\delta _{B_{a}}A_{b}^{2}%
}{4\pi S_{D}b^{2}\ln ^{2}(r_{H}/a)}\sum_{l=0}^{\infty }D_{l}\frac{\lambda
_{l}^{2}b^{2}\beta _{b}^{2}+4\zeta \beta _{b}-1}{\bar{I}_{0}^{(b)2}(\lambda
_{l}b)}.  \label{pbnearhor}
\end{eqnarray}%
In this limit the interaction forces have different signs for the Dirichlet
and non-Dirichlet boundary conditions on the brane $\xi =a$. The combination
$\delta _{B_{a}}p_{\mathrm{(int)}}^{(j)}$ is positive for large values $%
\beta _{b}$ and is negative for small values of this parameter. In the limit
$b\rightarrow \infty $ for fixed $a$, the dominant contribution comes from
the lowest mode $l=0$ and assuming that $A_{b}\neq \pm \lambda _{0}B_{b}$,
we have the estimates%
\begin{eqnarray}
p_{\mathrm{(int)}}^{(a)} &\approx &\frac{A_{a}^{2}e^{-2\lambda _{0}b}}{%
2S_{D}a^{2}r_{H}^{D-1}}\frac{A_{b}-\lambda _{0}B_{b}}{A_{b}+\lambda _{0}B_{b}%
}\int_{0}^{\infty }d\omega \,\frac{\left( \lambda _{l}^{2}a^{2}+\omega
^{2}\right) \beta _{a}^{2}+4\zeta \beta _{a}-1}{\bar{K}_{0}^{(a)2}(\lambda
_{0}a)},  \label{pafar} \\
p_{\mathrm{(int)}}^{(b)} &\approx &-\frac{\lambda _{0}e^{-2\lambda _{0}b}}{%
S_{D}br_{H}^{D-1}}\frac{A_{b}-\lambda _{0}B_{b}}{A_{b}+\lambda _{0}B_{b}}%
\int_{0}^{\infty }d\omega \frac{\bar{I}_{\omega }^{(a)}(\lambda _{0}a)}{\bar{%
K}_{\omega }^{(a)}(\lambda _{0}a)},  \label{pbfar}
\end{eqnarray}%
with the exponentially small interaction forces for both branes. In
particular, the combination $(A_{b}^{2}-\lambda _{0}^{2}B_{b}^{2})p_{\mathrm{%
(int)}}^{(j)}$ is positive/negative for large/small values $\beta
_{a}$. For small values of the curvature radius $r_{H}$, in the
way similar to that used for the VEV of the field square, it can
be seen that for a non-minimally coupled scalar field the main
contribution comes from the $l=0$ term and to the leading order we
have
\begin{equation}
p_{\mathrm{(int)}}^{(j)}\approx -\frac{\delta _{B_{a}}\delta _{B_{b}}[\zeta
n(n+1)]^{3/4}\sqrt{ab}}{2\sqrt{\pi }S_{D}r_{H}^{D+1/2}j\sqrt{b-a}}\exp [-2%
\sqrt{\zeta n(n+1)}(b-a)/r_{H}].  \label{pintlargecurv}
\end{equation}%
For a minimally coupled scalar field the dominant contribution comes from
the $l=0$ term with $\lambda _{0}=m$ and the interaction forces per unit
surface behave like $r_{H}^{1-D}$. In the near-horizon limit, $a,b\ll r_{H}$%
, in (\ref{pint2}) we replace the summation over $l$ by the integration in
accordance with (\ref{sumtoint}) and to the leading order the result for the
geometry of two parallel plates uniformly accelerated through the
Fulling-Rindler vacuum is obtained.

\section{Conclusion}

\label{sec:Conc}

In this paper, we investigate the polarization of the scalar vacuum induced
by two spherical branes in the $(D+1)$-dimensional bulk $Ri\times S^{D-1}$,
assuming that on the branes the field obeys the Robin boundary conditions.
In the corresponding braneworld scenario based on the orbifolded version of
the model the coefficients in the boundary conditions are expressed in terms
of the brane mass parameters by formula (\ref{ABbraneworld}). The most
important characteristics of the vacuum properties are the expectation
values of quantities bilinear in the field operator such as the field square
and the energy-momentum tensor. As the first step in the investigation of
these VEVs we evaluate the positive frequency Wightman function. The
corresponding mode sum contains the summation over the eigenfrequencies. In
the region between the branes the latter are the zeros of the bilinear
combination of the modified Bessel functions and their derivatives. For the
summation of the series over these zeros we employ a variant of the
generalized Abel-Plana formula. This allows us to present the Wightman
function as the sum of a single brane and second brane induced parts,
formulae (\ref{Wigh3}) and (\ref{Wigh31}).

The corresponding VEVs of the field square and the energy-momentum tensor
are obtained from the Wightman function in the coincidence limit and are
investigated in section \ref{sec:VEVEMT}. These VEVs are given by formulae (%
\ref{phi2sq1}) and (\ref{Tik1}), where $j=a,b$ provide two equivalent
representations. We have considered various limiting cases of the general
formulae. In particular, we have shown that when the left brane tends to the
horizon the interference parts in the VEVs of the field square and the
energy-momentum tensor vanish as $1/\ln ^{2}(r_{H}/a)$. In the limit when
the right brane tends to infinity, $b\rightarrow \infty $, the interference
parts are suppressed by the factor $\exp (-2\lambda _{0}b)$. In the high
curvature regime, corresponding to small values $r_{H}$, the behavior of the
VEVs is essentially different for minimally and non-minimally coupled scalar
fields. For a non-minimally coupled field the VEVs are suppressed by the
factor $\exp [-2\sqrt{\zeta n(n+1)}(\xi -a)/r_{H}]$ for single brane parts
and by $\exp [-2\sqrt{\zeta n(n+1)}(b-a)/r_{H}]$ for the interference parts.
For a minimally coupled field the main contribution comes from the $l=0$
term and the VEVs behave as $r_{H}^{1-D}$. In the limit when the both branes
are near the horizon, to the leading order the VEVs are obtained for the
geometry of two parallel plates uniformly accelerated through the
Fulling-Rindler vacuum.

In section \ref{sec:IntForce} we have investigated the vacuum forces acting
on the branes. These forces are presented as the sum of self-action and
interaction parts. Due to the well-known surface divergences, the
self-action part needs additional subtractions. The interaction forces are
finite for all nonzero interbrane distances and are given by formula (\ref%
{pint2}) or equivalently by (\ref{pint3}). In general, these forces are
different for the left and right branes and, in dependence of the values of
the parameters, they can be either attractive or repulsive. In particular,
at small interbrane distances they are repulsive for the Dirichlet boundary
condition on one brane and non-Dirichlet boundary condition on the other,
and are attractive for all other cases. When the left brane tends to the
horizon the interaction forces acting on the left and right branes behave as
$1/[a^{2}\ln (r_{H}/a)]$ and $1/\ln ^{2}(r_{H}/a)$, respectively. In the
limit when the $\xi $-coordinate of the right brane tends to infinity, the
interaction forces for both branes are suppressed by the factor $\exp
(-2\lambda _{0}b)$.

\section*{Acknowledgement}

AAS was supported in part by PVE/CAPES Program and by the Armenian Ministry
of Education and Science Grant No. 0124.

\end{document}